\documentclass[aps,prl,twocolumn,superscriptaddress,balancelastpage,showpacs,reprint]{revtex4-2}

\usepackage{blindtext}
\usepackage{centernot}
\usepackage{graphicx}
\graphicspath{{img}}
\usepackage{amsmath,bbold}
\usepackage{times}
\usepackage{amssymb}
\usepackage{mathrsfs}
\usepackage{chemarr}
\usepackage{color}
\usepackage{url}
\usepackage{version}
\usepackage[hidelinks]{hyperref}
\usepackage{mwe,tikz}
\usepackage[percent]{overpic}
\usepackage{bm}
\usepackage[export]{adjustbox}
\definecolor{linkcolor}{rgb}{0,0,0.6} 
\usepackage{lipsum}

\usepackage{dcolumn}

\usepackage[utf8]{inputenc}
\usepackage[T1]{fontenc}
\usepackage{mathptmx}
\usepackage{stmaryrd}
\usepackage{ulem} 
\usepackage{xcolor}
\usepackage{textcomp}
\usepackage{gensymb}
\usepackage{siunitx} 

\newcommand{\um}{\,\si{\micro\meter}}

\newcommand{\MSD}{\text{MSD} }

\begin{document}

\title{Evolving motility of active droplets is captured by a self-repelling random walk model}


\author{Wenjun Chen}
\affiliation{\mbox{Center for Soft Matter Research, Physics Department, New York University, New York, New York 10003, USA}}

\author{Adrien Izzet}
\affiliation{\mbox{Center for Soft Matter Research, Physics Department, New York University, New York, New York 10003, USA}}
\affiliation{\mbox{Université Paris-Saclay, INRAE, AgroParisTech, UMR SayFood, 91120 Palaiseau, France}}

\author{Ruben Zakine}%
\affiliation{\mbox{Center for Soft Matter Research, Physics Department, New York University, New York, New York 10003, USA}}
\affiliation{Courant Institute, New York University, 251 Mercer Street, New York, New York 10012, USA}

\affiliation{LadHyX UMR CNRS 7646, École Polytechnique, 91128 Palaiseau Cedex, France}

\author{Eric Cl\'ement}%
\affiliation{\mbox{Laboratoire PMMH-ESPCI Paris, PSL  University, Sorbonne University, 7, Quai Saint-Bernard, 75005 Paris, France}}
\affiliation{Institut Universitaire de France (IUF), 75005 Paris, France}

\author{Eric Vanden-Eijnden}%
\affiliation{Courant Institute, New York University, 251 Mercer Street, New York, New York 10012, USA}

\author{Jasna Brujic}
\email{jb2929@nyu.edu}
\affiliation{\mbox{Center for Soft Matter Research, Physics Department, New York University, New York, New York 10003, USA}}
\affiliation{\mbox{Laboratoire PMMH-ESPCI Paris, PSL  University, Sorbonne University, 7, Quai Saint-Bernard, 75005 Paris, France}}

\date{\today}

\begin{abstract}
	Swimming droplets are a class of active particles whose motility changes as a function of time due to shrinkage and self-avoidance of their trail. Here we combine experiments and theory to show that our non-Markovian droplet (NMD) model, akin to a true self-avoiding walk~\cite{amit1983asymptotic}, quantitatively captures droplet motion. We thus estimate the effective temperature arising from hydrodynamic flows and the coupling strength of the propulsion force as a function of fuel concentration. This framework explains a broad range of phenomena, including memory effects, solute-mediated interactions, droplet hovering above the surface, and enhanced collective diffusion.

\end{abstract}

\maketitle

Active droplets are a class of artificial microswimmers that consume energy from the environment to fuel their motility~\cite{maass2016swimming, moran2017phoretic, lohse2020physicochemical, michelin2023self}. Changing the temperature~\cite{kruger2016curling, ramesh2023arrested}, viscosity~\cite{hokmabad2021emergence, dwivedi2023mode}, droplet size~\cite{suga2018self,suda2021straight, ray2023experimental}, fuel concentration~\cite{izzet2020tunable}, or geometric confinement~\cite{dey2022oscillatory}, has been shown to lead to different types of droplet motility. These experimental swimming trajectories range from straight and helical to random and oscillatory, owing to solute-mediated interactions.
Typically, a swimming droplet dissolves into micelles and leaves an oil trail behind, which then creates a repulsive concentration gradient that results in self-avoidance~\cite{schulz2005feedback, hokmabad2022chemotactic, jin2017chemotaxis, moerman2017solute, izzet2020tunable}. The droplet can cross its chemical trail with a finite probability determined by the associated energy cost. As a result, the mean square displacement of these droplets cannot be described by Flory's self-avoiding polymer theory~\cite{flory1953principles} but rather falls into a class of self-interacting random walks~\cite{derkachov_renormalisability_1989, grassberger_2017, barbier-chebbah_self-interacting_2022}.

Theoretical models that couple chemotactic motion with the diffusing chemical field secreted by the swimmers have been developed~\cite{tsori2004self, grima2005strong, grima2006phase, sengupta2009dynamics, taktikos2011modeling, kranz2016effective, daftari2022self}. They display features such as
self-trapping due to chemoattractant clouds~\cite{tsori2004self}, enhanced diffusion~\cite{grima2005strong, grima2006phase}, and behavior akin to Active Brownian Particles (ABP)~\cite{sengupta2009dynamics, daftari2022self}, emphasizing the nuanced interplay of chemical fields and swimmer motion. To describe experimental trajectories, previous works have either used the ABP model at short times~\cite{howse2007self, volpe2011microswimmers, izri2014self, izzet2020tunable} or taken into account local chemical gradients to capture collision interactions and trail persistence times~\cite{hokmabad2022chemotactic, jin2017chemotaxis}.

\begin{figure*}
	\centering
	\includegraphics[width=1\textwidth]{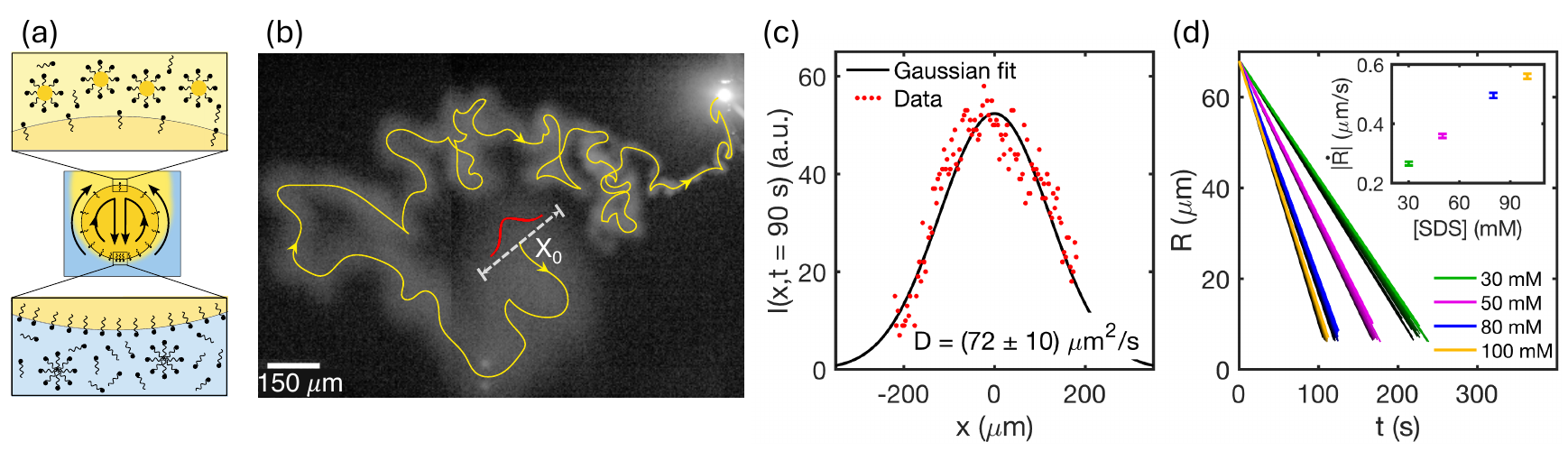}
	\caption{(a) Schematic of the droplet propulsion mechanism via interfacial Marangoni gradients. (b) A snapshot at the end of a 90s trajectory of a swimming droplet shows the diffusive trail of fluorescent oily micelles ($R_0= \SI{35}{\um}$, [SDS] $= \SI{30}{\milli M}$). (c) A Gaussian fit using Eq.~\eqref{eq:Itx} of the fluorescence intensity (red in b) yields the micellar diffusion constant $D = $ \SI[separate-uncertainty = true]{72(10)}{\um^2/s}. (d) The droplet radius decreases linearly with time, giving a solubilization rate that increases as a function of SDS concentration in the inset. }
	\label{fig:trail}
\end{figure*}

Here we combine experiments and theory to characterize the motion of a single droplet over the course of its lifetime. Our coarse-grained model captures droplet shrinkage through the use of a mollified delta function~\cite{daftari2022self, daftari2024memory}, which avoids the problem of infinitely strong self-interactions arising with a point source approximation~\cite{jin2017chemotaxis, hokmabad2022chemotactic}. Moreover, we take into account the full trail of the droplet to model self-avoidance rather than only local chemical gradients~\cite{jin2017chemotaxis, jin2019fine, hokmabad2022chemotactic}. Our model not only describes the evolving motility over time, but it also quantitatively estimates the self-propulsion force, the level of noise in the system through an effective temperature $T_{\text{eff}}$, as well as the coupling strength $\beta$ between the droplet and its chemical gradient.

We find that $T_{\text{eff}}$ is orders of magnitude higher than thermal energy at room temperature, suggesting that the noise arises from hydrodynamic flows surrounding the droplet. Indeed, chemo-advection mechanisms give rise to fluid flows, which can in turn randomly change the direction of droplet motion~\cite{michelin2013spontaneous, morozov2019self, morozov2019nonlinear, morozov2020adsorption, hu2022spontaneous}. As a function of surfactant concentration, our results show that the effective temperature decreases while the coupling strength $\beta$ remains constant. In quantifying the chemical gradient force, we discover that it is strong enough to overcome gravity, and thus explain the fact that the droplet hovers significantly above the surface. Finally, we extend the motility study to many-body effects and show that the collective behavior strongly depends on the history of the sample.

Our experimental system consists of a single droplet of diethyl phthalate (DEP, Sigma) with an initial radius of around \SI{80}{\um}, which is injected into an aqueous solution of sodium dodecyl sulphate (SDS)~\cite{izzet2020tunable}.
As shown in Fig.~\ref{fig:trail}a, a droplet immersed in an SDS solution above
the critical micelle concentration (CMC) starts dissolving and forms oily micelles homogeneously around the droplet. This symmetry can be spontaneously broken due to fluctuations in the system. Given the fact that there is a difference in the surfactant CMC in DEP-saturated water and pure water, the asymmetry creates a surface tension gradient, giving rise to a Marangoni flow that drives self-sustained droplet motion. The droplet therefore leaves a trail of oily micelles as it moves. A snapshot at the end of the trajectory ($t=90$ s) reveals the evolution of the diffusive trail over time, as shown in Fig.~\ref{fig:trail}b.  This trail is fluorescently labeled using Nile Red dye, which preferentially dissolves in the oil.

In Fig.~\ref{fig:trail}c, we show the fluorescence intensity profile at the starting point $\mathbf{X}_0$ of the trajectory.
To model trail diffusion, we approximate
the droplet as a 3D Gaussian sphere emitting oil, $G_R(\mathbf{x}) = \frac{1}{(2\pi R^2)^{3/2}} \exp\left(-\frac{|\mathbf{x}|^2}{2R^2}\right)$, yielding the density profile at position $\mathbf{x}$ around the emission site $\mathbf{X}_0$,
\begin{equation}
	\label{eq:Itx}
	c_t(\mathbf{x}) = \frac{c_0}{(2\pi(R^2 + 2Dt))^{3/2}} \exp\left(-\frac{|\mathbf{x}-\mathbf{X}_0|^2}{2(R^2 + 2Dt)}\right),
\end{equation}
where the initial standard deviation equals the droplet radius $R$, $D$ is the diffusion constant of the filled micelles, and $t$ is the diffusion time. The intensity in Fig.~\ref{fig:trail}c is directly proportional to the concentration field. Fitting the data to Eq.~\eqref{eq:Itx} yields $D = $ \SI[separate-uncertainty = true]{72(10)}{\um^2/s}, which corresponds to the radius of filled micelles \SI[separate-uncertainty = true]{3.4(0.5)}{nm}, consistent with values reported in the literature \cite{missel1980thermodynamic, hiemenz2016principles}. Measuring droplet radius over time (Fig.~\ref{fig:trail}d) shows a linear decrease:  $R_t = R_0 - |\dot R| t$,
where $|\dot R|=cst$ is the shrinkage rate of the droplet radius. The shrinkage rate is related to the droplet oil emission rate $ \alpha_t\simeq \frac{3R_t^2}{{\delta^3}}|\dot R| $, where $\delta$ is the size of the oil sphere in the micelle,  approximated by the micellar radius. With increasing SDS concentration the shrinkage rate increases, as shown in the inset.
\begin{figure*}
	\centering
	\includegraphics[width=0.9\textwidth]{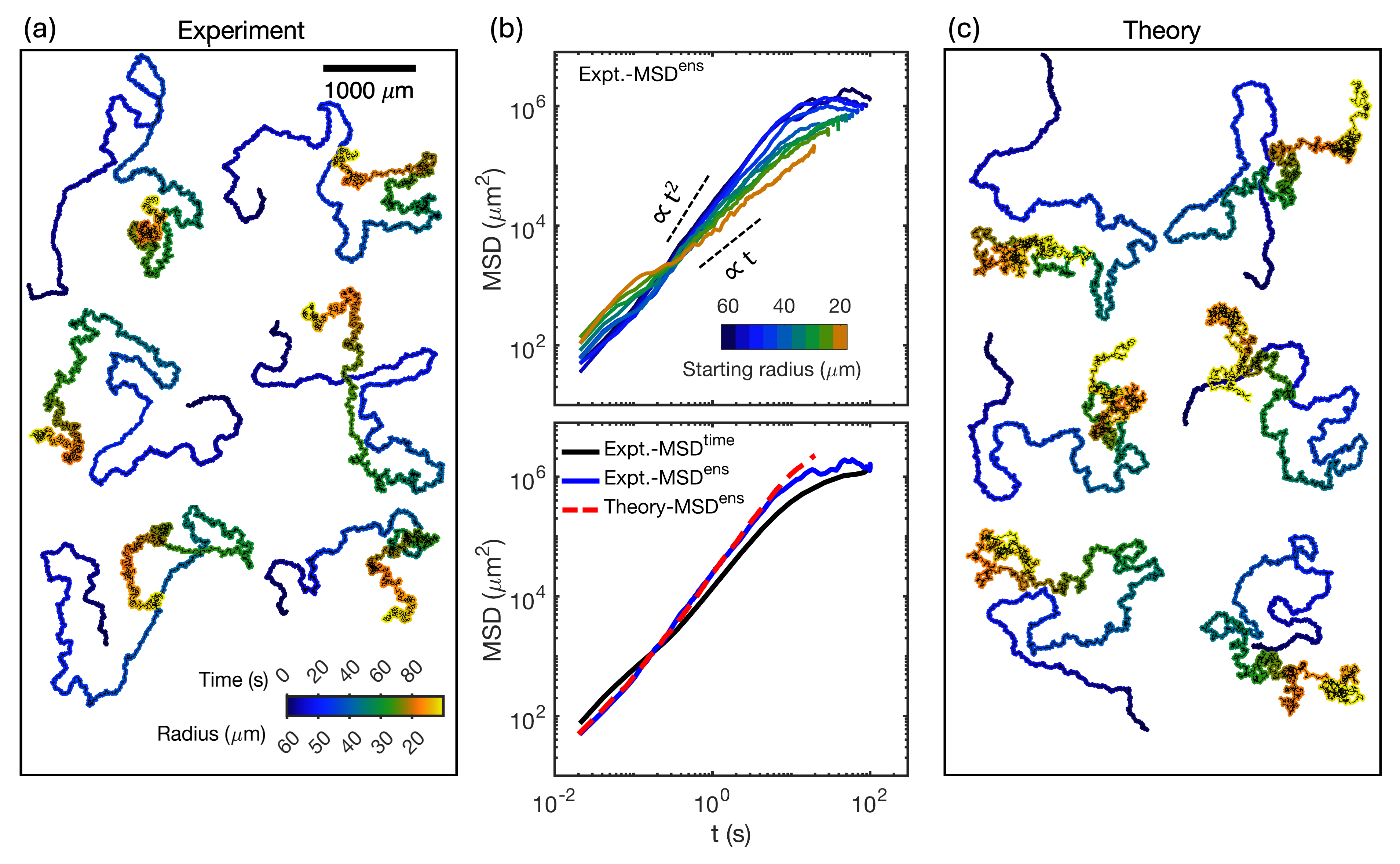}
	\caption{(a) Experimental trajectories at [SDS] $= \SI{80}{\milli M}$ that evolve with time,  displaying self-avoidance, crossing, and increasing noise. (b) Top: MSD\textsuperscript{ens} of experimental data show dependence on the size range of the shrinking droplet. Bottom: MSD\textsuperscript{ens} (blue line) at a starting radius \SI{60}{\um} shows a discrepancy from the MSD\textsuperscript{time} (black line), indicating memory effects. The model fits the data very well with \(\beta = 0.038 \, k_B T \mathrm{M^{-1}}\), where $\mathrm M\equiv \mathrm{mol/L}$ and \(T_{eff} = 8 \times 10^4 T_{bath}\). Note that the model does not capture the plateau in the data at long times due to sample-cell confinement or pollution. (c) Model-generated trajectories display the same features as experimental ones shown in (a).}
	\label{fig:traj}
\end{figure*}

At [SDS] $= \SI{80}{\milli M}$, Fig.~\ref{fig:traj}a,b shows single droplet trajectories that evolve from super-diffusive to diffusive behavior on intermediate timescales, as the droplet shrinks.
To quantify the time evolution of the trajectories we compare the ensemble-averaged mean squared displacement (MSD) for different sections of the trajectory (Fig.~\ref{fig:traj}b):
\begin{equation}
	\label{eq:msd:ens}
	\MSD^{\text{ens}}_t= \frac{1}{N} \sum_{i=1}^{N} \left|\mathbf{X}^{(i)}_{t+t_0} - \mathbf{X}^{(i)}_{t_0}\right|^2,
\end{equation}
where \(t_0\) is the starting time and $N$ is the number of trajectories indexed by $i$. The top panel shows the change in the MSD as a function of the initial radius at which the section begins. As a result of this non-Markovian behavior, the bottom panel shows a discrepancy between the ensemble-averaged and the time-averaged MSD:
\begin{equation}
	\label{eq:msd:time}
	\MSD^{\text{time}}_t = \frac{1}{N(M_t+1)} \sum_{i=1}^{N} \sum_{j=0}^{M_t} \left| \mathbf{X}^{(i)}_{t+j \Delta t} - \mathbf{X}^{(i)}_{j \Delta t} \right|^2,
\end{equation}
where $\tau$ is the time length of each trajectory,  \({\Delta} t\) $= \SI{0.02}{s}$, and $M_t = \lfloor (\tau-t)/\Delta t\rfloor$.

To explain these phenomena, we employ a non-Markovian droplet model (NMD) in which complex hydrodynamics and chemo-advection mechanisms are represented as a fundamental coupling between chemical gradients and droplet motion. In this model, the time-evolving oil field $c_t(\mathbf{x})$ is governed by a 3D diffusion equation,
while the droplet motion is described by a stochastic differential equation with a constant noise amplitude $\sigma$ corresponding to the hydrodynamic fluctuations generated by the dissolving droplet:
\begin{subequations}
	\renewcommand{\theequation}{\theparentequation\alph{equation}}
	\begin{align}
		\label{eq:ctx}
		\partial_t c_t(\mathbf{x})  & = D\nabla^2 c_t(\mathbf{x}) + \alpha_t G_{R_t} \left( \mathbf{x} - \mathbf{X}_t \right),                                                                                      \\
		\label{eq:Xt}
		\gamma_t \dot{\mathbf{X}}_t & = -\beta \frac{R_t^3}{{\delta}^3} \int_{\mathbb{R}^3} G_{R_t} \left( \mathbf{x} - \mathbf{X}_t \right) \nabla c_t(\mathbf{x}) \, d\mathbf{x} + \sqrt{{\sigma}}\mathbf{\xi}_t.
	\end{align}
\end{subequations}
The integral is performed in 3D, using the reflecting boundary condition modeled by an image droplet on the bottom of the sample cell (see Supplemental Material\cite{SM}). Eq.~\eqref{eq:ctx} has a diffusion term employing the experimentally determined \(D\) of the oily micelles and
a source term that depends on the instantaneous position \( \mathbf{X}_t \) of the moving droplet emitting oil at a rate of $\alpha_t$.
Eq.~\eqref{eq:Xt} is a modified Langevin equation for a Brownian particle in a force field with the inertial term neglected in the low-Reynolds number regime. The friction coefficient is given by \({\gamma_t} = 6{\pi}{\eta}R_t\), where \({\eta}\) is the viscosity of the aqueous phase. The droplet response to the concentration gradient is defined as a convolution of the concentration gradient with the mollified delta function, multiplied by \( {\beta} R_t^3/{\delta}^3 \), with \( {\beta} \) as the coupling strength, and \(R_t^3/{\delta}^3\) denoting the amount of oil carried by the droplet. Note that the force depends on both the gradient of the oil field and \(R_t^3/{\delta}^3\).
From the hydrodynamic point of view, the force is generated by the Marangoni flow and has therefore been proposed to scale with the droplet surface~\cite{daftari2022self}. However, the droplet emits the oil volumetrically from the interior and the response to the Marangoni forcing occurs inside and outside the droplet~\cite{morozov2019nonlinear, schmitt2013swimming, schmitt2016marangoni}, such that the propulsion force depends on $R^3$.

To eliminate the oil field, we can integrate Eq.~\eqref{eq:ctx} to obtain
\begin{equation}
	\label{eq:ctx:s}
	c_t(\mathbf{x}) = \int_0^t \frac{\alpha_s\exp\left(-\frac{|\mathbf{x}-\mathbf{X}_s|^2}{2(R_s^2 + 2D(t-s))}\right)}{(2\pi(R_s^2 + 2D(t-s))^{3/2}}ds.
\end{equation}
Note that Eq.~\eqref{eq:Itx} is an approximation of Eq.~\eqref{eq:ctx:s} that is valid if the droplet moves fast, and if we look at positions $\mathbf{x}$ close to the initial $\mathbf{X}_0$.
Inserting Eq.~\eqref{eq:ctx:s} in Eq.~\eqref{eq:Xt} and performing the Gaussian integral over $\mathbf{x}$ we get a closed non-Markovian equation for the position of the droplet
\begin{equation}
	\label{eq:X:NMM}
	\begin{aligned}
		 & \dot{\mathbf{X}}_t = \omega^2 \int_0^t r_t^2r_s^2 \frac{\mathbf{X}_t - \mathbf{X}_s}{\left( r_t^2/2 + r_s^2/2 + D(t-s)/R_0^2 \right)^{5/2}} \\
		 & \times \exp\left(-\frac{|\mathbf{X}_t - \mathbf{X}_s|^2}{2R_0^2\left( r_t^2 + r_s^2 + 2D(t-s)/R_0^2 \right)}\right) ds
		+ \frac{\sqrt{\sigma}}{\gamma_0 r_t}\xi_t,
	\end{aligned}
\end{equation}
where $\gamma_0 = 6{\pi}{\eta}R_0$, $\omega^2 = \frac{3}{16{\pi}^{3/2}{\gamma_0}{\delta}^6}|\dot R|{\beta}$, and $r_t = 1 - |\dot R|t/R_0$. The first term on the right-hand side, which represents the drift velocity $v$, is determined by an integral over all of its past, incorporating memory effects.  The known or experimentally determined input parameters are $D = $ \SI{72}{\um^2/s}, $\delta = $ \SI{3.4}{nm}, initial size \(R_0 = \) \SI{80}{\um}, \({\eta} = \) \SI{0.89}{mPa \cdot s}, and the shrinkage rate $|\dot R| =\SI{0.5}{\um/s}$. This leaves two fitting parameters, the coupling strength $\beta$ and the noise amplitude $\sigma$ (see Supplemental Material\cite{SM}). 

\begin{figure}
	\centering
	\includegraphics[width=0.48\textwidth]{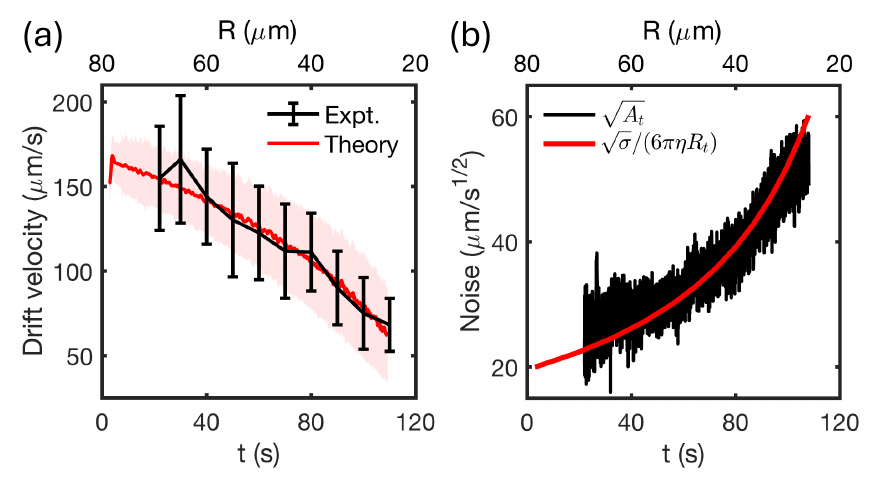}
	\caption{ As the droplet shrinks, the experimental decrease in drift velocity in (a) and the increase in the noise amplitude $\sqrt{A_t}$ in (b) (See Supplemental Material\cite{SM}) are captured by the respective terms on the right-hand-side of Eq.(6) using parameters from the MSD fit in Fig.~\ref{fig:traj}b. The errors in (a) come from averaging over $\sim$100 and 1000 individual trajectories for experiments and theory, respectively, while the fluctuations in (b) arise from measurement error.}
	\label{fig:velocity}
\end{figure}
\begin{figure*}
	\centering
	\includegraphics[width=0.9\textwidth]{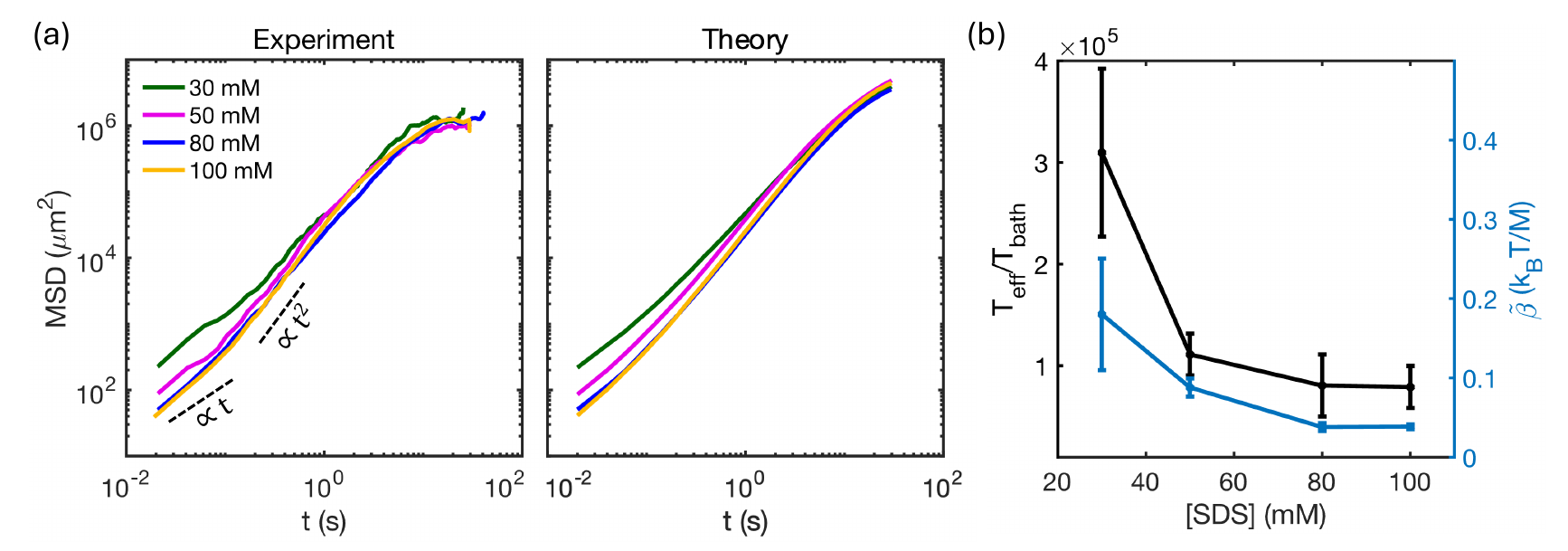}
	\caption{(a) MSD\textsuperscript{ens} with a starting droplet size of \SI{60}{\um} show good agreement between the model and the data as a function of [SDS]. They exhibit a fast diffusive, then ballistic, and finally a slow diffusive regime. In the experiment, the MSD saturates at long times due to the confinement of the motion by the sample cell. (b) Fitting parameters $T_\mathrm{eff}$ (black) and coupling $ \tilde{\beta}=\beta\delta_0^6/\delta^6$
		(blue) decrease as a function of [SDS], but the known increase in $\delta$ with [SDS]\cite{pisarvcik2015determination} quantitatively matches the trend in $\tilde\beta$, implying a constant coupling strength $\beta$. }
	\label{fig:parameter}
\end{figure*}

As shown in Fig.~\ref{fig:traj}b (bottom panel), the model fits the ensemble-averaged MSD of the data very well. Furthermore, trajectories generated with the model reproduce experimental features: self-avoidance and the motility evolution with time. More specifically, we show quantitative agreement between theory and experiments in terms of the drift velocity (i.e. propulsion force) and the noise term, as shown in Fig.~\ref{fig:velocity}. The experimental drift velocity~$v$ is obtained by trajectory smoothing in a regime where the result is independent of the smoothing frequency and the experimental noise amplitude, $\sqrt{A_t}$, is calculated over the given frame rate (see Supplemental Material\cite{SM}). Note that characterizing the trajectory with its P\'eclet number defined by the instantaneous velocity ($\text{Pe} = |\dot X_t| R_t/D$) is intrinsically frame rate-dependent. This is how the P\'eclet number is defined in some previous studies~\cite{hokmabad2021emergence, suda2021straight, hu2019chaotic, ramesh2023arrested}, in which case the P\'eclet number increases along the trajectory. As droplets shrink over time, here we show that the drift velocity $v$ decreases, while the noise term dramatically increases. The decreasing $v$ can be rationalized by the reduced emission rate at the droplet interface, while the increasing noise amplitude
is a direct result of droplet shrinkage.
The excellent agreement between the data and the model {\it a posteriori} justifies the assumed dependence of chemical response on the amount of oil carried by the droplet $R_t^3/{\delta}^3$, and the assumption of a constant noise amplitude $\sigma$.

\begin{figure*}
	\centering
	\includegraphics[width=1\textwidth]{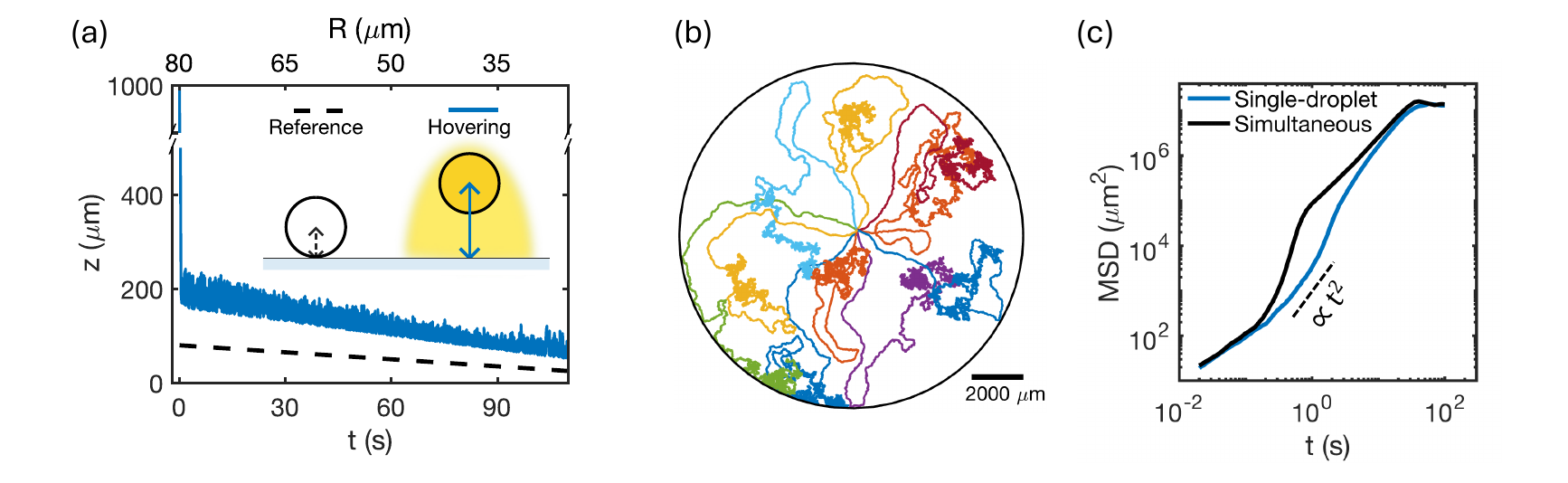}
	\caption{Extensions to the model using parameters from the MSD fit in Fig.~\ref{fig:traj}b.
		(a) A simulated active droplet is injected into the sample cell at a height \(z = \SI{1000}{\um}\) and rapidly falls to its equilibrium hovering height around \(z = \SI{200}{\um}\), where the propulsion force balances gravity. The dashed line shows the shrinking radius on the surface for reference. (b) Multiple droplets are injected simultaneously in the center of the sample cell. (c) These droplets exhibit an enhanced MSD\textsuperscript{ens} compared to that of a single droplet (starting size $= \SI{80}{\um}$). Their motion is accelerated due to their repulsive interactions at the center. }
	\label{fig:collective}
\end{figure*}

Next, we vary the SDS concentration to study the effect of shrinkage rate on droplet motility, as shown in Fig. 1 of the Supplemental Material\cite{SM}. Previously, we showed that small droplets of \SI{15}{\um} follow the ABP model at short times with a decrease in persistence time with [SDS], but the large droplets considered here undergo a different motility type.
Qualitatively, the trajectories become shorter and less noisy as a function of [SDS]. This indicates that their drift velocity and effective temperature decrease. Quantitatively, the MSD of our mesoscopic model is in good agreement with experimental data, as shown in Fig.~\ref{fig:parameter}a. As [SDS] increases, the short-time diffusion in the MSD is suppressed, while the ballistic regime exhibits a decrease in the drift velocity. This observation appears counterintuitive: at higher [SDS], oily micelles form more rapidly causing an increased rate of energy transfer to the droplets, which should lead to faster swimming speeds. However, we speculate that the fast oil solubilization into micelles surrounding the droplets occurs in all directions, which leads to weaker surface tension gradients at the interface, reducing the motility speed. Fits to the data give estimates for the noise amplitude \({\sigma}\) and the coupling \( \tilde{\beta}\) as a function of SDS concentration.

Assuming the noise \({\sigma}\) has the same form as white noise \(2{\gamma_0}k_B T_{\text{eff}}\), with \(k_B \) as the Boltzmann constant, and \(T_{\text{eff}}\) as the effective temperature, we find the \(T_{\text{eff}}\) is \(10^5\) times the bath temperature (Fig.~\ref{fig:parameter}b), indicating that thermal fluctuations are not at the origin of the noise of the swimmer. As [SDS] increases, both the effective temperature and the instantaneous velocity decrease (see Fig.~3 of Supplemental Material\cite{SM}). This result is consistent with hydrodynamic analyses as a function of P\'eclet number \cite{hu2019chaotic, morozov2019nonlinear, hokmabad2021emergence, hu2022spontaneous}. With increasing P\'eclet number the droplets switch more frequently between dipolar and quadrupolar modes, resulting in more chaotic behavior, i.e., higher effective temperatures.

The chemical coupling \( \tilde{\beta}\) decreases with increasing [SDS] concentration, as shown in Fig.~\ref{fig:parameter}b. Since $ \tilde{\beta}={\beta}\delta_0^6/\delta^6 $, where ${\delta_0}$ is a constant for dimensionless normalization, this decrease can be attributed to the known increase in the micellar size up to $\sim1.3\delta$ over our experimental range of [SDS]~\cite{bezzobotnov1988temperature, christov2004synergistic, pisarvcik2015determination}. This trend quantitatively explains our results, implying a constant coupling strength \(\beta = 0.2 \, k_B T \mathrm{M^{-1}}\). Explicitly counting the number of oily micelles in our model allows for this estimate, which is important in understanding chemotaxis \cite{jin2017chemotaxis}.

One manifestation of the coupling strength is that an active droplet, which is so far shown to be propelled in the $x,y$ plane, is also lifted vertically (in $z$) against gravity by the self-propulsion force above the surface boundary, as shown in Fig.~\ref{fig:collective}a. As the droplet exudes a spherically symmetric concentration gradient, the presence of the surface boundary breaks this symmetry propelling the droplet upwards until the gradient force matches the downward buoyancy force ($\sim 1 \mathrm{nN}$). The equilibrium height evolves with droplet shrinkage, in agreement with our previous experimental observations \cite{moerman2017solute} and theoretical work \cite{desai2021instability}.

Building on the insights from single droplets, our model is readily extended to explore collective effects theoretically. As shown in Fig.~\ref{fig:collective}b, ten droplets with an initial size of \SI{80}{\um} are injected into the center of the sample cell simultaneously (see movies in the Supplemental Material\cite{SM}). Distinct from the single-droplet scenario, these droplets feel the repulsion from not only their own trails but also from their neighbors' trails. As a result, the MSD is significantly enhanced, giving rise to a super-ballistic regime (see Fig.~\ref{fig:collective}c). This result highlights the fact that collective phenomena must include the spatio-temporal history of droplet creation.

In the future, our NMD model can be readily applied to other microswimmers, such as Janus particles or living microorganisms, to better characterize their chemotaxis in terms of the relevant physical control parameters. The underlying chemo-advection mechanisms need further elaboration using a more complex hydrodynamic analysis.

The authors thank S\'ebastien Michelin, Sascha Hilgenfeldt, Stefano Martiniani, Katherine A. Newhall, Arnold Mathijssen, and John Brady for their enlightening discussions.
EVE is supported by the NSF  DMR-1420073, DMS-2012510, and DMS-2134216, by the Simons Collaboration on Wave Turbulence, Grant No. 617006, and by a Vannevar Bush Faculty Fellowship.
JB iss supported by the NSF DMR grant No. 2105255 and the Swiss National Science Foundation through Grant No. 10000141.

\bibliographystyle{apsrev4-2}
\bibliography{swimmer, SM}

\end{document}


\preprint{APS/123-QED}

\title{Supplemental material for ``Evolving motility of active droplets is captured by a self-repelling random walk model''}

\author{Wenjun Chen}
\affiliation{\mbox{Center for Soft Matter Research, Physics Department, New York University, New York, New York 10003, USA}}

\author{Adrien Izzet}
\affiliation{\mbox{Center for Soft Matter Research, Physics Department, New York University, New York, New York 10003, USA}}
\affiliation{\mbox{Université Paris-Saclay, INRAE, AgroParisTech, UMR SayFood, 91120 Palaiseau, France}}

\author{Ruben Zakine}%
\affiliation{\mbox{Center for Soft Matter Research, Physics Department, New York University, New York, New York 10003, USA}}
\affiliation{Courant Institute, New York University, 251 Mercer Street, New York, New York 10012, USA}
\affiliation{LadHyX UMR CNRS 7646, École Polytechnique, 91128 Palaiseau Cedex, France}

\author{Eric Cl\'ement}%
\affiliation{\mbox{Laboratoire PMMH-ESPCI Paris, PSL  University, Sorbonne University, 7, Quai Saint-Bernard, 75005 Paris, France}} 
\affiliation{Institut Universitaire de France (IUF), 75005 Paris, France}

\author{Eric Vanden-Eijnden}%
\affiliation{Courant Institute, New York University, 251 Mercer Street, New York, New York 10012, USA}

\author{Jasna Brujic}
\email{jb2929@nyu.edu}
\affiliation{\mbox{Center for Soft Matter Research, Physics Department, New York University, New York, New York 10003, USA}}


\maketitle

\subsection{Experimental methods}
Approximately \SI{2}{nL} of Diethyl phthalate (DEP, Sigma) oil is injected into a circular sample cell (1 cm wide and 1 mm deep) filled with SDS solution at the desired concentration, via a custom-built microfluidic setup controlled by intracellular microinjection dispense systems (Picospritzer III). This setup generates droplets with an approximate radius of \SI{80}{\um}. Droplet trajectories are observed using a Nikon Eclipse Ti microscope equipped with a 4 \(\times\) Plan Apochromat $\lambda$ Nikon objective (NA = 0.20). Movies are recorded using a Zyla sCMOS Andor camera operating at 50 frames/s.


\subsection{Data analysis}

In the absence of an analytic formula for the mean squared displacement (MSD) of Non-Markovian Droplet (NMD) model, we fit the data by generating $1000$ trajectories using the model with the free parameters which are iteratively tuned to achieve optimal agreement with the data. 

To determine the drift velocity, the trajectories are sliced into 10-second segments, which are then smoothed using a moving average filter. With an increasing smoothing window size, the velocity reaches a plateau, which corresponds to the drift velocity (see Fig.~\ref{fig:smoothing}). The error bar comes from N \(\sim\) 100 trajectories.

We use the following equation to extract the noise from experimental trajectories, 
\begin{equation}
\label{eq:noise}
A_t = \frac{\frac{1}{N} \sum_{i=1}^{N} \left|\mathbf{X}^{(i)}_{t+\Delta t} - \mathbf{X}^{(i)}_{t}\right|^2 - v^2 \Delta t^2}{2 \Delta t}, 
\end{equation}
where $\Delta t = 0.02 \mathrm{s}$, and $v$ is the drift velocity, and N is the total number of trajectories. The data shown in Fig.3b of the main text shows \(\sqrt{A_t}\).

\subsection{Non-Markovian Droplet (NMD) model}
Here we derive Eq. 6 of the main text in more detail. As mentioned in the manuscript:
\begin{subequations}
\renewcommand{\theequation}{\theparentequation\alph{equation}}
\begin{align}
\label{eq:ctx}
\partial_t c_t(\mathbf{x})&= D\nabla^2 c_t(\mathbf{x}) + \alpha_t G_{R_t} \left( \mathbf{x} - \mathbf{X}_t \right),\\
\label{eq:Xt}
\gamma_t \dot{\mathbf{X}}_t &= -\beta \frac{R_t^3}{{\delta}^3} \int_{\mathbb{R}^3} G_{R_t} \left( \mathbf{x} - \mathbf{X}_t \right) \nabla c_t(\mathbf{x}) \, d\mathbf{x} + \sqrt{{\sigma}}\mathbf{\xi}_t.
\end{align}
\end{subequations}
where \(R_t = R_0 - |\dot R| t\), \({\alpha_t} = \frac{3R_t^2}{{\delta}^3}|\dot R|\), \({\gamma_t}=6{\pi\eta}R_t\), and \(G_R(\mathbf{x}) = \frac{1}{(2\pi R^2)^{3/2}} \exp\left(-\frac{|\mathbf{x}|^2}{2R^2}\right)\).

Starting from the equation describing the chemical field, and taking its Fourier transform gives:
\begin{equation}
\partial_t \hat{c}_t(\mathbf{k}) = -D|\mathbf{k}|^2 \hat{c}_t(\mathbf{k}) + \alpha_t e^{-\frac{1}{2}k^2 R_t^2 + i\mathbf{k}\cdot \mathbf{X}_t}.
\end{equation}

The solution is then given by:
\begin{equation}
\hat{c}_t(\mathbf{k}) = \int_{0}^{t} ds \, e^{-k^2 D(t-s)} \alpha_s e^{-\frac{1}{2}k^2 R_s^2 + i\mathbf{k}\cdot \mathbf{X}_s},
\end{equation}
where \(k^2 = |\mathbf{k}|^2\).

\begin{align}
c_t(\mathbf{x}) &= \frac{1}{(2\pi)^3} \int d\mathbf{k} \, e^{-i\mathbf{k}\cdot \mathbf{x}} \hat{c}_t(\mathbf{k}) \notag \\
&= \frac{1}{(2\pi)^3} \int_0^t ds \, \alpha_s \int d\mathbf{k} \, e^{-k^2D(t-s)} e^{-\frac{1}{2}k^2R_s^2} e^{-i\mathbf{k}\cdot (\mathbf{x}-\mathbf{X}_s)} \notag \\
&= \frac{1}{(2\pi)^{3/2}} \int_0^t ds \, \alpha_s \frac{1}{\left( 2D(t-s)+R_s^2 \right)^{3/2}} e^{-\frac{(\mathbf{x}-\mathbf{X}_s)^2}{2(2D(t-s)+R_s^2)}}.
\end{align}


On the other hand, the Gaussian kernel can be expressed as: 
\begin{align}
G_R(\mathbf{x}) & = \frac{1}{(2\pi R^2)^{3/2}} e^{-\frac{|\mathbf{x}|^2}{2R^2}} \notag \ \\
& = \frac{1}{(2\pi)^3} \int d\mathbf{k} e^{-\frac{1}{2}k^2R^2+i\mathbf{k}\cdot \mathbf{x}}
\end{align}

Consequently, the Langevin equation is given by:

\begin{align}
\gamma_t \dot{\mathbf{X}}_t &= -\beta \frac{R_t^3}{{\delta}^3} \int d\mathbf{x} G_{R_t}(\mathbf{x} - \mathbf{X}_t) \nabla c_t(\mathbf{x})  + \sqrt{{\sigma}}\mathbf{\xi}_t \notag \ \\
&= -\beta \frac{R_t^3}{{\delta}^3} \int d\mathbf{x} \frac{1}{(2\pi)^3} \int d\mathbf{k} e^{-\frac{1}{2}k^2R^2+i\mathbf{k}\cdot(\mathbf{x}-\mathbf{X}_t)} \nabla c_t(\mathbf{x})  + \sqrt{{\sigma}}\mathbf{\xi}_t \notag \ \ \\
&= -\beta \frac{R^3_t}{{\delta}^3} \int d\mathbf{k} (-i\mathbf{k}) \frac{1}{(2\pi)^3} e^{-\frac{1}{2}k^2R_t^2-i\mathbf{k}\cdot \mathbf{X}_t} \int d\mathbf{x} c_t(\mathbf{x})e^{i\mathbf{k}\cdot \mathbf{x}}  + \sqrt{{\sigma}}\mathbf{\xi}_t \notag \ \\
&= -\beta \frac{R_t^3}{{\delta}^3} \int d\mathbf{k} (-i\mathbf{k}) \frac{1}{(2\pi)^3} e^{-\frac{1}{2}k^2R_t^2-i\mathbf{k}\cdot \mathbf{X}_t} \hat{c}_t(\mathbf{k})  + \sqrt{{\sigma}}\mathbf{\xi}_t \notag \\
&= -\beta \frac{R_t^3}{{\delta}^3} \int d\mathbf{k} (-\mathbf{k}) \frac{1}{(2\pi)^3} e^{-\frac{1}{2}k^2R_t^2-i\mathbf{k}\cdot \mathbf{X}_t} \int_0^t ds \, e^{-k^2D(t-s)} {\alpha}_s e^{-\frac{1}{2}k^2 R_s^2+i\mathbf{k}\cdot \mathbf{X}_s} + \sqrt{{\sigma}}\mathbf{\xi}_t \notag \\ 
&=  \beta \frac{R_t^3}{(2\pi)^{3/2}{\delta}^3} \int_0^t ds \, {\alpha}_s \frac{\mathbf{X}_t - \mathbf{X}_s}{(R_t^2 + R_s^2 + 2D(t - s))^{5/2}} e^{-\frac{|\mathbf{X}_t-\mathbf{X}_s|^2}{2(R_t^2+R_s^2+2D(t-s))}}  + \sqrt{{\sigma}}\mathbf{\xi}_t,  
\end{align}
Therefore we get a closed non-Markovian equation for the position of the droplet (Eq. 6 in the main text):
\begin{equation}
\dot{\mathbf{X}}_t = \omega^2 \int_0^t r_t^2r_s^2 \frac{\mathbf{X}_t - \mathbf{X}_s}{\left( r_t^2/2 + r_s^2/2 + D(t-s)/R_0^2 \right)^{5/2}} \\
\exp\left(-\frac{|\mathbf{X}_t - \mathbf{X}_s|^2}{2R_0^2\left( r_t^2 + r_s^2 + 2D(t-s)/R_0^2 \right)}\right) ds \\
+ \frac{\sqrt{\sigma}}{\gamma_0 r_t}\xi_t,
\end{equation}
where \({\gamma_0} = 6{\pi}{\eta}R_0\), \({\omega}^2 = \frac{3}{16{\pi}^{3/2}{\gamma_0}{\delta}^6}\frac{dR}{dt}{\beta}\), and \(r_t = 1 - |\dot R| t/R_0\).

\subsection{Reflecting boundary condition}
In the experiments, the droplet sinks over the surface located at $X = (x, y, z)$. Such a boundary effect can be seen as a repulsive interaction with a mirror droplet located at $X' = (x, y, -z)$, therefore, the propulsion force is not only arising from its own trail, given by Eq.~(6) of the main text, but also from its image droplet. The deterministic part of the equation is then given by: 
\begin{equation}
\label{eq:hovering}
\begin{split}
\dot{\mathbf{X}}_t =& \omega^2 \int_0^t r_t^2r_s^2 \frac{\mathbf{X}_t - \mathbf{X}_s}{\left( r_t^2/2 + r_s^2/2 + D(t-s)/R_0^2 \right)^{5/2}} 
\exp\left(-\frac{|\mathbf{X}_t - \mathbf{X}_s|^2}{2R_0^2\left( r_t^2 + r_s^2 + 2D(t-s)/R_0^2 \right)}\right) ds \\
&+ \omega^2 \int_0^t r_t^2r_s^2 \frac{\mathbf{X}_t - \mathbf{X'}_s}{\left( r_t^2/2 + r_s^2/2 + D(t-s)/R_0^2 \right)^{5/2}} \exp\left(-\frac{|\mathbf{X}_t - \mathbf{X'}_s|^2}{2R_0^2\left( r_t^2 + r_s^2 + 2D(t-s)/R_0^2 \right)}\right) ds \\ 
&- f_g\mathbf{e_z},
\end{split}
\end{equation}
where \(- f_g\mathbf{e_z}\) is the buoyancy force.

\subsection{Collective behavior}
Our model is readily applied to study collective behaviour of droplets by taking into account the propulsion from the trail of all the droplets. The position of droplet j is denoted by \(\mathbf{X}^{(j)}_t\), where \(j = 1,2,..,N\), with N representing the total number of droplets in the sample cell. The closed non-Markovian equation for the position of the droplet j is given by:
\begin{equation}
\dot{\mathbf{X}}^{(j)}_t = \omega^2 \sum_{i=1}^{N} \int_0^t r_t^2r_s^2 \frac{\mathbf{X}^{(j)}_t - \mathbf{X}^{(i)}_s}{\left( r_t^2/2 + r_s^2/2 + D(t-s)/R_0^2 \right)^{5/2}} \\
\exp\left(-\frac{|\mathbf{X}^{(j)}_t - \mathbf{X}^{(i)}_s|^2}{2R_0^2\left( r_t^2 + r_s^2 + 2D(t-s)/R_0^2 \right)}\right) ds \\
+ \frac{\sqrt{\sigma}}{\gamma_0 r_t}\xi^{(j)}_t.
\end{equation}

\begin{figure}[H]
 \centering
 \includegraphics[width=0.9\textwidth]{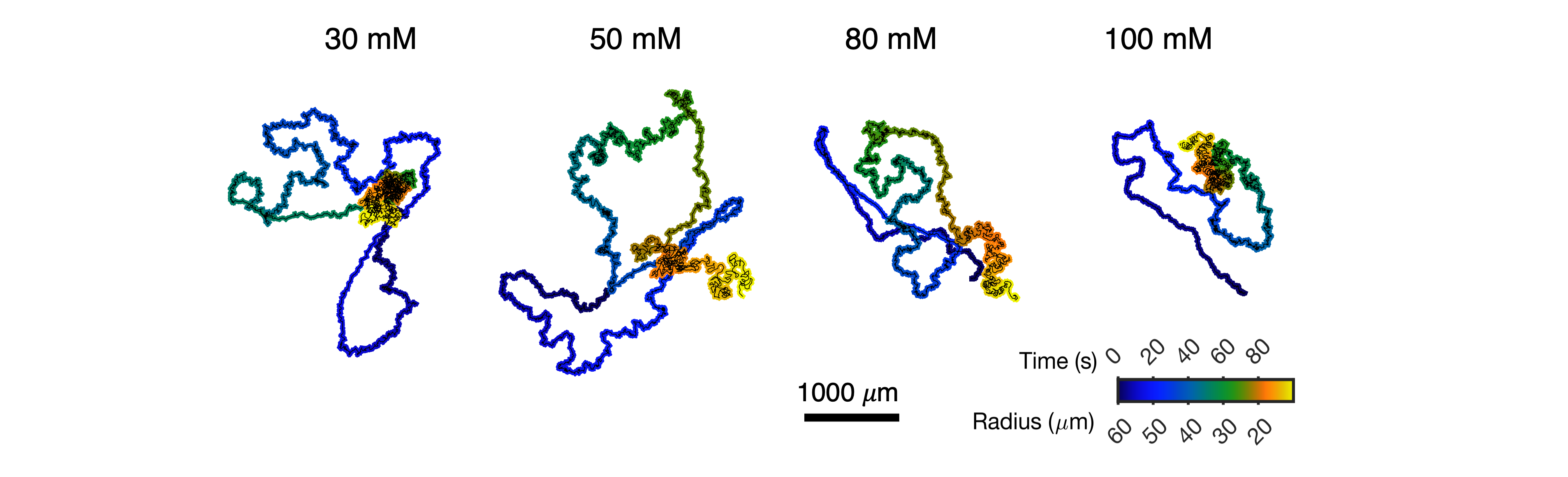}
 \caption{Droplets at different [SDS] display similar features: self-avoidance behavior, crossing, and motility evolution.} With in increase [SDS], the contour length of the trajectories decreases, indicating a smaller drift velocity.
 \label{fig:traj_SDS}
\end{figure}

\begin{figure}[H]
 \centering
 \includegraphics[width=1\textwidth]{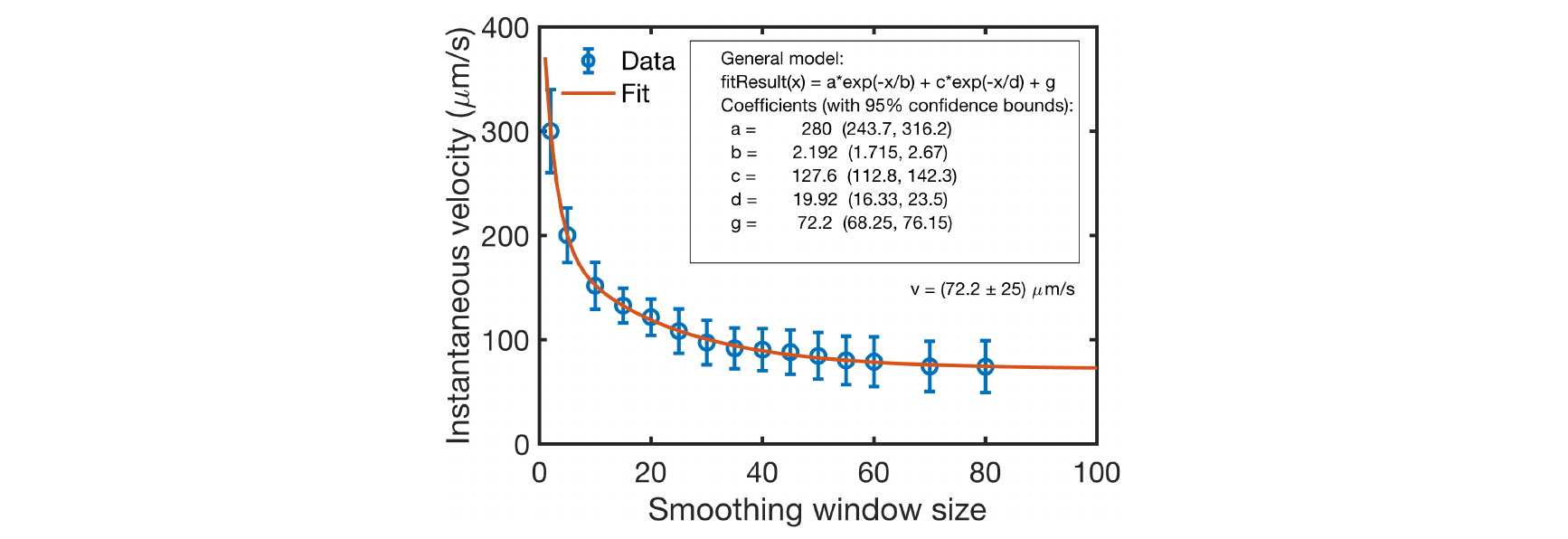}
 \caption{An example of trajectory smoothing: [SDS] $= \SI{80}{\milli M}$, initial droplet size of \SI{30}{\um}, and a segment length of \SI{10}{s}. Each data point represents the instantaneous velocity averaged over the ensemble of the smoothed trajectories. As the smoothing window size increases, the velocity reaches a plateau value at \SI{72}{\um/s}.}
 \label{fig:smoothing}
\end{figure}

\begin{figure}[H]
 \centering
 \includegraphics[width=1\textwidth]{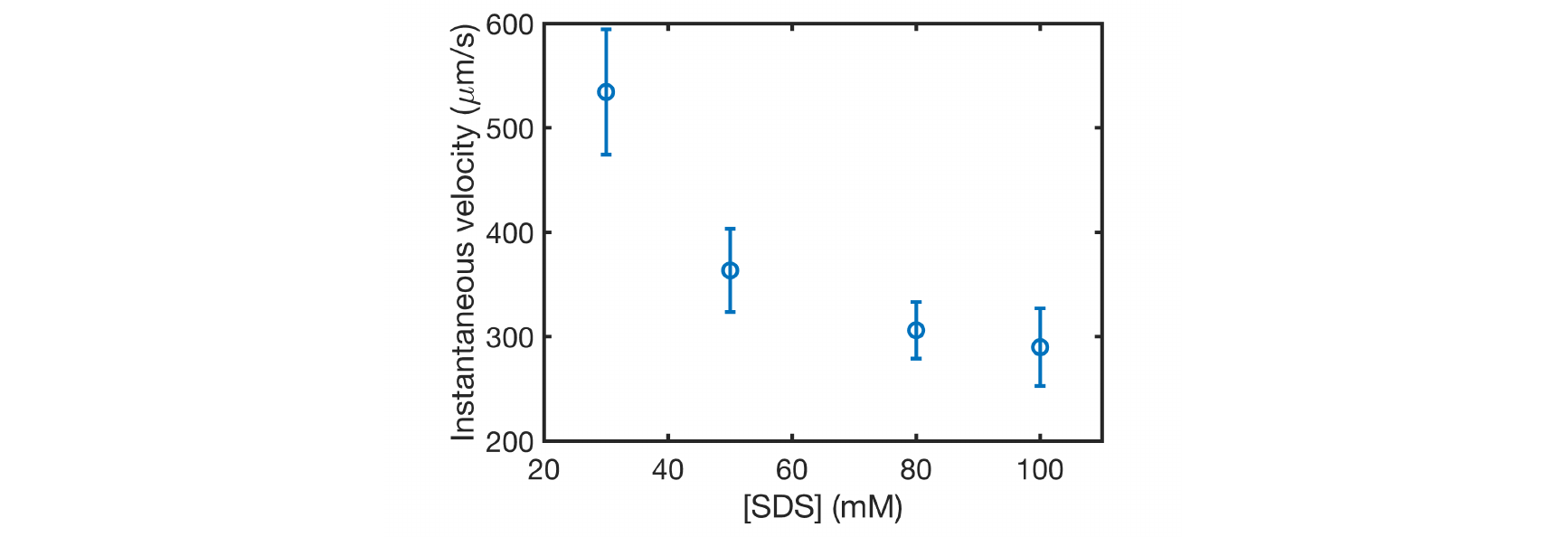}
 \caption{For droplets with a starting size of \SI{60}{\um}, with an increasing [SDS], the instantaneous velocity decreases, corresponding to a lower P\'eclet number.}
 \label{fig:instant_v}
\end{figure}